\documentclass[10pt,conference]{IEEEtran}
\IEEEoverridecommandlockouts
\usepackage[T1]{fontenc}
\usepackage{cite}
\usepackage{amsmath}
\usepackage{graphicx}
\usepackage{booktabs}
\usepackage{array}
\usepackage{multirow}
\usepackage{url}
\usepackage{xspace}

\newcommand{\SolvesZeroFiveBQthree}{220}
\newcommand{\PassZeroFiveBQthree}{40.6}

\newcommand{\SolvesZeroFiveBQfour}{229}
\newcommand{\PassZeroFiveBQfour}{42.3}

\newcommand{\SolvesZeroFiveBQfive}{232}
\newcommand{\PassZeroFiveBQfive}{42.8}

\newcommand{\PassZeroFiveBQsix}{44.5}

\newcommand{\PassZeroFiveBQeight}{45.2}

\newcommand{\PassZeroFiveBFpsixteen}{45.6}

\newcommand{\SolvesOneFiveBQthree}{303}
\newcommand{\PassOneFiveBQthree}{55.9}

\newcommand{\SolvesOneFiveBQfour}{316}
\newcommand{\PassOneFiveBQfour}{58.3}

\newcommand{\PassOneFiveBQfive}{57.9}

\newcommand{\PassOneFiveBQsix}{58.1}

\newcommand{\PassOneFiveBQeight}{57.7}

\newcommand{\PassOneFiveBFpsixteen}{58.1}

\newcommand{\SolvesThreeBQthree}{7}
\newcommand{\PassThreeBQthree}{1.3}

\newcommand{\SolvesThreeBQfour}{366}
\newcommand{\PassThreeBQfour}{67.5}

\newcommand{\SolvesThreeBQfive}{367}
\newcommand{\PassThreeBQfive}{67.7}

\newcommand{\PassThreeBQsix}{66.1}

\newcommand{\PassThreeBQeight}{67.0}

\newcommand{\PassThreeBFpsixteen}{66.2}

\newcommand{\SolvesSevenBQtwo}{363}
\newcommand{\PassSevenBQtwo}{67.0}

\newcommand{\SolvesSevenBQthree}{388}
\newcommand{\PassSevenBQthree}{71.6}

\newcommand{\SolvesSevenBQfour}{396}
\newcommand{\PassSevenBQfour}{73.1}

\newcommand{\PassSevenBQfive}{72.9}

\newcommand{\SolvesSevenBQsix}{403}
\newcommand{\PassSevenBQsix}{74.4}

\newcommand{\PassSevenBQeight}{73.8}

\newcommand{\SerialChangedOutcomes}{9}
\newcommand{\SerialChangedConfigs}{8}

\newcommand{\SerialHeChanges}{0}

\newcommand{\AdjDiffSevenBQtwoQthree}{+4.6}

\newcommand{\AdjPSevenBQtwoQthree}{0.011}

\newcommand{\AdjPHolmSevenBQtwoQthree}{0.053}
\newcommand{\AdjCILoSevenBQtwoQthree}{+1.3}
\newcommand{\AdjCIHiSevenBQtwoQthree}{+7.9}
\newcommand{\AdjSigHolmSevenBQtwoQthree}{no}

\newcommand{\EndDiffZeroFiveB}{+5.0}

\newcommand{\EndPHolmZeroFiveB}{0.006}

\newcommand{\EndCILoZeroFiveB}{+1.8}
\newcommand{\EndCIHiZeroFiveB}{+8.1}
\newcommand{\EndSigHolmZeroFiveB}{yes}
\newcommand{\EndDiffOneFiveB}{+2.2}

\newcommand{\EndPHolmOneFiveB}{0.224}

\newcommand{\EndCILoOneFiveB}{$-$1.1}
\newcommand{\EndCIHiOneFiveB}{+5.5}
\newcommand{\EndSigHolmOneFiveB}{no}
\newcommand{\EndDiffThreeB}{+64.9}

\newcommand{\EndPHolmThreeB}{$<$0.001}

\newcommand{\EndCILoThreeB}{+60.9}
\newcommand{\EndCIHiThreeB}{+69.0}
\newcommand{\EndSigHolmThreeB}{yes}
\newcommand{\EndDiffSevenB}{+6.8}

\newcommand{\EndPHolmSevenB}{$<$0.001}

\newcommand{\EndCILoSevenB}{+3.5}
\newcommand{\EndCIHiSevenB}{+10.3}
\newcommand{\EndSigHolmSevenB}{yes}
\newcommand{\QfourTopDiffZeroFiveB}{+3.3}

\newcommand{\QfourTopPZeroFiveB}{0.036}
\newcommand{\QfourTopPHolmZeroFiveB}{0.142}

\newcommand{\QfourTopCINinetyLoZeroFiveB}{+0.9}
\newcommand{\QfourTopCINinetyHiZeroFiveB}{+5.7}

\newcommand{\QfourTopVerdictZeroFiveB}{non-inferiority not shown}

\newcommand{\QfourTopDiffOneFiveB}{$-$0.2}

\newcommand{\QfourTopPHolmOneFiveB}{1.000}

\newcommand{\QfourTopCINinetyLoOneFiveB}{$-$2.6}
\newcommand{\QfourTopCINinetyHiOneFiveB}{+2.2}

\newcommand{\QfourTopVerdictOneFiveB}{equivalent}

\newcommand{\QfourTopDiffThreeB}{$-$1.3}

\newcommand{\QfourTopPHolmThreeB}{0.972}

\newcommand{\QfourTopCINinetyLoThreeB}{$-$3.1}
\newcommand{\QfourTopCINinetyHiThreeB}{+0.6}

\newcommand{\QfourTopVerdictThreeB}{non-inferior}
\newcommand{\QfourTopQfourBetterUpToThreeB}{3.1}
\newcommand{\QfourTopQfourWorseUpToThreeB}{0.6}

\newcommand{\QfourTopDiffSevenB}{+0.7}

\newcommand{\QfourTopPHolmSevenB}{1.000}

\newcommand{\QfourTopCINinetyLoSevenB}{$-$1.1}
\newcommand{\QfourTopCINinetyHiSevenB}{+2.6}

\newcommand{\QfourTopVerdictSevenB}{equivalent}

\newcommand{\EquivMarginPP}{3}
\newcommand{\HeadSolvesA}{366}
\newcommand{\HeadSolvesB}{363}

\newcommand{\HeadP}{0.836}
\newcommand{\HeadDiffPP}{+0.55}
\newcommand{\HeadCILo}{$-$2.95}
\newcommand{\HeadCIHi}{+4.06}
\newcommand{\HeadCINinetyLo}{$-$2.40}
\newcommand{\HeadCINinetyHi}{+3.51}
\newcommand{\HeadMaxDeficit}{2.4}

\newcommand{\HeadOverlapBoth}{318}
\newcommand{\HeadOverlapOnlyA}{48}
\newcommand{\HeadOverlapOnlyB}{45}
\newcommand{\HeadOverlapNeither}{131}

\newcommand{\SizeGainThreeSevenMin}{5.2}
\newcommand{\SizeGainThreeSevenMax}{8.3}

\newcommand{\AnchorHeSevenBQeight}{84.1}
\newcommand{\AnchorMbppSevenBQeight}{69.3}

\newcommand{\IdleRailsWThree}{0.509}

\newcommand{\JpsZeroFiveBQthree}{38.9}

\newcommand{\MjTokZeroFiveBQthree}{79}

\newcommand{\JpsZeroFiveBQfour}{42.3}

\newcommand{\MjTokZeroFiveBQfour}{93}

\newcommand{\JpsZeroFiveBQfive}{41.3}

\newcommand{\MjTokZeroFiveBQfive}{94}

\newcommand{\JpsZeroFiveBQsix}{49.1}

\newcommand{\MjTokZeroFiveBQsix}{115}

\newcommand{\JpsZeroFiveBQeight}{43.1}

\newcommand{\MjTokZeroFiveBQeight}{103}

\newcommand{\JpsZeroFiveBFpsixteen}{46.0}

\newcommand{\MjTokZeroFiveBFpsixteen}{110}

\newcommand{\JpsOneFiveBQthree}{53.4}

\newcommand{\MjTokOneFiveBQthree}{179}

\newcommand{\JpsOneFiveBQfour}{39.3}

\newcommand{\MjTokOneFiveBQfour}{157}

\newcommand{\JpsOneFiveBQfive}{57.2}

\newcommand{\MjTokOneFiveBQfive}{236}

\newcommand{\JpsOneFiveBQsix}{44.9}

\newcommand{\MjTokOneFiveBQsix}{185}

\newcommand{\JpsOneFiveBQeight}{49.5}

\newcommand{\MjTokOneFiveBQeight}{203}

\newcommand{\JpsOneFiveBFpsixteen}{69.1}

\newcommand{\MjTokOneFiveBFpsixteen}{284}

\newcommand{\JpsThreeBQthree}{4{,}708}

\newcommand{\MjTokThreeBQthree}{317}

\newcommand{\JpsThreeBQfour}{69.4}
\newcommand{\JpsCILoThreeBQfour}{64}
\newcommand{\JpsCIHiThreeBQfour}{75}

\newcommand{\MjTokThreeBQfour}{294}

\newcommand{\JpsThreeBQfive}{101.2}

\newcommand{\MjTokThreeBQfive}{423}

\newcommand{\JpsThreeBQsix}{86.4}

\newcommand{\MjTokThreeBQsix}{352}

\newcommand{\JpsThreeBQeight}{89.1}

\newcommand{\MjTokThreeBQeight}{365}

\newcommand{\JpsThreeBFpsixteen}{116.1}

\newcommand{\MjTokThreeBFpsixteen}{470}

\newcommand{\JpsSevenBQtwo}{141.2}
\newcommand{\JpsCILoSevenBQtwo}{131}
\newcommand{\JpsCIHiSevenBQtwo}{153}

\newcommand{\MjTokSevenBQtwo}{585}

\newcommand{\JpsSevenBQthree}{187.0}

\newcommand{\MjTokSevenBQthree}{848}

\newcommand{\JpsSevenBQfour}{145.7}

\newcommand{\MjTokSevenBQfour}{663}

\newcommand{\JpsSevenBQfive}{215.1}

\newcommand{\MjTokSevenBQfive}{963}

\newcommand{\JpsSevenBQsix}{183.1}

\newcommand{\MjTokSevenBQsix}{837}

\newcommand{\JpsSevenBQeight}{187.1}

\newcommand{\MjTokSevenBQeight}{858}

\newcommand{\HeadJpsA}{69.4}
\newcommand{\HeadJpsB}{141.2}
\newcommand{\HeadJpsRatio}{2.03}

\newcommand{\HeadJpsRatioCILo}{1.91}
\newcommand{\HeadJpsRatioCIHi}{2.17}
\newcommand{\HeadJpsRatioHe}{1.95}

\newcommand{\HeadJpsRatioMbpp}{2.11}

\newcommand{\HeadJpsRatioClean}{2.03}

\newcommand{\HeadJpsRatioSysLo}{1.66}
\newcommand{\HeadJpsRatioSysHi}{1.81}

\newcommand{\JpsBrokenThreeB}{4{,}708}
\newcommand{\BrokenJpsRatio}{68}
\newcommand{\BrokenRankTok}{14}
\newcommand{\BrokenRankJps}{24}
\newcommand{\CleanIdleJpsMin}{1.7}
\newcommand{\CleanIdleJpsMax}{4.0}
\newcommand{\RailsShareMin}{28}
\newcommand{\RailsShareMax}{53}
\newcommand{\FrontierList}{0.5B-\texttt{q3\_K\_M}, 1.5B-\texttt{q4\_K\_M}, 3B-\texttt{q4\_K\_M}, 3B-\texttt{q5\_K\_M}, 7B-\texttt{q4\_K\_M}, 7B-\texttt{q6\_K}}
\newcommand{\FrontierN}{6}

\newcommand{\FrontierNSys}{8}
\newcommand{\MarginZeroFiveBQthreeOn}{1.2}
\newcommand{\MarginSevenBQeightOff}{2.2}
\newcommand{\GapSevenBQfourOverQtwo}{3.2}

\newcommand{\MarginFpOffOneFiveBCount}{43}

\newcommand{\MarginFpOffOneFiveBSysLo}{39}
\newcommand{\MarginFpOffOneFiveBSysHi}{41}
\newcommand{\MarginFpOffThreeBCount}{40}

\newcommand{\MarginFpOffThreeBSysLo}{35}
\newcommand{\MarginFpOffThreeBSysHi}{41}

\newcommand{\FrontierMarginOneFiveBQthree}{26.3}
\newcommand{\FrontierMarginOneFiveBQfour}{76.4}

\newcommand{\FrontierMarginThreeBQfour}{45.8}

\newcommand{\FrontierMarginSevenBQtwo}{50.9}
\newcommand{\FrontierMarginSevenBQthree}{22.0}

\newcommand{\ScaleRatioQthree}{0.04}
\newcommand{\ScaleRatioQfour}{2.10}

\newcommand{\ScaleRatioSixFour}{2.64}
\newcommand{\ScaleGainFour}{5.5}
\newcommand{\ScaleGainSixFour}{6.8}
\newcommand{\ScaleRatioMin}{2.10}
\newcommand{\ScaleRatioMax}{2.13}
\newcommand{\CheapestGapZeroFiveB}{6.3}
\newcommand{\CheapestGapOneFiveB}{14.1}
\newcommand{\CheapestGapThreeB}{24.5}
\newcommand{\CheapestGapSevenB}{3.2}

\newcommand{\RepeatIdentical}{164}
\newcommand{\RepeatDecodePct}{+23.6}
\newcommand{\RepeatWallPct}{$-$21.0}

\newcommand{\BrokenRepeatIdentical}{50}

\newcommand{\BestValueTop}{7B-\texttt{q6\_K}}
\newcommand{\BestValueScorePct}{90.8}
\newcommand{\BestValueEnergyPct}{38}

\newcommand{\FileBytesGBThreeBQfour}{1.93}
\newcommand{\FileBytesGBSevenBQtwo}{3.02}
\newcommand{\TensorQfourZeroFiveB}{12}
\newcommand{\TensorQuantZeroFiveB}{169}
\newcommand{\TensorQsixZeroFiveB}{24}
\newcommand{\BrokenBlob}{9bb5d28b}
\newcommand{\BrokenManifest}{65ff2bc1}
\newcommand{\BpwZeroFiveBQthree}{5.66}

\newcommand{\ListedGBZeroFiveBQthree}{0.36}

\newcommand{\BpwZeroFiveBQfour}{6.35}

\newcommand{\ListedGBZeroFiveBQfour}{0.40}

\newcommand{\BpwZeroFiveBQfive}{6.71}

\newcommand{\ListedGBZeroFiveBQfive}{0.42}

\newcommand{\BpwZeroFiveBQsix}{8.09}

\newcommand{\ListedGBZeroFiveBQsix}{0.51}

\newcommand{\BpwZeroFiveBQeight}{8.50}

\newcommand{\ListedGBZeroFiveBQeight}{0.53}

\newcommand{\BpwZeroFiveBFpsixteen}{16.00}

\newcommand{\ListedGBZeroFiveBFpsixteen}{0.99}

\newcommand{\BpwOneFiveBQthree}{4.24}

\newcommand{\ListedGBOneFiveBQthree}{0.82}

\newcommand{\BpwOneFiveBQfour}{5.08}

\newcommand{\ListedGBOneFiveBQfour}{0.99}

\newcommand{\BpwOneFiveBQfive}{5.80}

\newcommand{\ListedGBOneFiveBQfive}{1.10}

\newcommand{\BpwOneFiveBQsix}{6.56}

\newcommand{\ListedGBOneFiveBQsix}{1.30}

\newcommand{\BpwOneFiveBQeight}{8.50}

\newcommand{\ListedGBOneFiveBQeight}{1.60}

\newcommand{\BpwOneFiveBFpsixteen}{16.00}

\newcommand{\ListedGBOneFiveBFpsixteen}{3.10}

\newcommand{\BpwThreeBQthree}{4.11}

\newcommand{\ListedGBThreeBQthree}{1.60}

\newcommand{\BpwThreeBQfour}{4.99}

\newcommand{\ListedGBThreeBQfour}{1.90}

\newcommand{\BpwThreeBQfive}{5.75}

\newcommand{\ListedGBThreeBQfive}{2.20}

\newcommand{\BpwThreeBQsix}{6.56}

\newcommand{\ListedGBThreeBQsix}{2.50}

\newcommand{\BpwThreeBQeight}{8.50}

\newcommand{\ListedGBThreeBQeight}{3.30}

\newcommand{\BpwThreeBFpsixteen}{16.00}

\newcommand{\ListedGBThreeBFpsixteen}{6.20}

\newcommand{\BpwSevenBQtwo}{3.16}

\newcommand{\ListedGBSevenBQtwo}{3.00}

\newcommand{\BpwSevenBQthree}{3.99}

\newcommand{\ListedGBSevenBQthree}{3.80}

\newcommand{\BpwSevenBQfour}{4.91}

\newcommand{\ListedGBSevenBQfour}{4.70}

\newcommand{\BpwSevenBQfive}{5.71}

\newcommand{\ListedGBSevenBQfive}{5.40}

\newcommand{\BpwSevenBQsix}{6.56}

\newcommand{\ListedGBSevenBQsix}{6.30}

\newcommand{\BpwSevenBQeight}{8.50}

\newcommand{\ListedGBSevenBQeight}{8.10}

\newcommand{\RuntimeOverheadMin}{2.7}
\newcommand{\RuntimeOverheadMax}{19.2}

\newcommand{\HeadListedLessPct}{36.7}

\newcommand{\HeadListedMorePct}{58}

\newcommand{\BudgetHalfPick}{0.5B-\texttt{q5\_K\_M}}

\newcommand{\BudgetOnePick}{1.5B-\texttt{q4\_K\_M}}

\newcommand{\BudgetTwoPick}{3B-\texttt{q4\_K\_M}}
\newcommand{\BudgetTwoPass}{67.5}

\newcommand{\BudgetTwoJps}{69}
\newcommand{\BudgetFourPick}{7B-\texttt{q3\_K\_M}}
\newcommand{\BudgetFourPass}{71.6}

\newcommand{\BudgetFourJps}{187}
\newcommand{\BudgetEightPick}{7B-\texttt{q6\_K}}

\newcommand{\BudgetOnePickRuntime}{1.5B-\texttt{q3\_K\_M}}
\newcommand{\BudgetTwoPickRuntime}{1.5B-\texttt{q4\_K\_M}}
\newcommand{\BudgetFourPickRuntime}{3B-\texttt{q5\_K\_M}}

\newcommand{\MarginalSolveA}{259}
\newcommand{\MarginalSolveB}{1{,}077}

\newcommand{\SessionTokSwing}{19.5}

\newcommand{\PickPHalf}{0.810}

\newcommand{\PickPOne}{0.203}

\newcommand{\PickPFour}{0.024}

\newcommand{\PickPEight}{0.248}

\newcommand{\IsoThreeBOverOneFive}{9.2}
\newcommand{\TtwoValidityCoveragePct}{98}

\newcommand{\TtwoRRailsThree}{1.925}

\newcommand{\TtwoRRailsLoThree}{1.718}
\newcommand{\TtwoRRailsHiThree}{2.157}

\newcommand{\TtwoRSys}{1.86}

\newcommand{\TtwoRSysLo}{1.66}
\newcommand{\TtwoRSysHi}{2.09}

\newcommand{\TtwoDecode}{2.03}

\newcommand{\TtwoDecodeLo}{2.00}
\newcommand{\TtwoDecodeHi}{2.06}

\newcommand{\TtwoWall}{1.98}

\newcommand{\TtwoWallLo}{1.92}
\newcommand{\TtwoWallHi}{2.03}

\newcommand{\TtwoRRailsNoPairOne}{2.005}
\newcommand{\TtwoRRailsNoPairOneLo}{1.971}
\newcommand{\TtwoRRailsNoPairOneHi}{2.041}

\newcommand{\TtwoPairsFlagged}{5}
\newcommand{\TtwoRRailsEndIdle}{1.929}
\newcommand{\TtwoRRailsAugustIdle}{1.926}
\newcommand{\TtwoAugustRRails}{1.956}
\newcommand{\TtwoHeavyMin}{90}
\newcommand{\TtwoHeavyMax}{97}
\newcommand{\TtwoValidBlocks}{10}

\newcommand{\JpsRatioTwoOverOne}{1.76}
\newcommand{\SessionTokSwingMedian}{3.7}

\newcommand{\SessionTokSwingFactor}{1.43}
\newcommand{\RuleMarginRatioMin}{1.44}
\newcommand{\RuleStableLo}{33}
\newcommand{\RuleStableHi}{40}
\newcommand{\TtwoBlockSwingMax}{1.24}

\newcommand{\NonInfPHolmOneFiveB}{0.041}

\newcommand{\NonInfPHolmThreeB}{$<$0.001}

\newcommand{\NonInfPHolmSevenB}{0.041}

\newcommand{\NonInfScoreHolmOneFiveB}{0.051}

\newcommand{\NonInfScoreHolmThreeB}{$<$0.001}

\newcommand{\NonInfScoreHolmSevenB}{0.051}
\newcommand{\AnchorNonInfScoreP}{0.024}
\newcommand{\AllPairsN}{276}
\newcommand{\AllPairsPTwo}{$<$0.001}
\newcommand{\AllPairsGainQthree}{1.000}
\newcommand{\AllPairsGainQfour}{0.103}
\newcommand{\AllPairsGainQsix}{0.010}
\newcommand{\AllPairsSevenBQfourQtwo}{0.031}
\newcommand{\HeadSolvesHeA}{134}
\newcommand{\HeadSolvesMbppA}{232}
\newcommand{\HeadSolvesHeB}{131}

\newcommand{\TtwoBelowMainPct}{1.5}
\newcommand{\TtwoAboveMainPct}{2.6}
\newcommand{\TtwoNoPairOneAB}{2.005}
\newcommand{\TtwoNoPairOneBA}{2.006}

\newcommand{\SevenBQsixLeadMax}{2.95}
\newcommand{\EquivMarginTasks}{16}
\newcommand{\EquivMarginStableLo}{2.6}
\newcommand{\EquivMarginStableHi}{3.1}
\newcommand{\GapWeightBoundPct}{1}
\newcommand{\NearPairThreeOnly}{48}
\newcommand{\NearPairSevenOnly}{48}
\newcommand{\NearPairP}{1.000}

\newcommand{\OneGbSystemLeadWord}{one}

\newcommand{\TokFrontierHalfBExtraWord}{four}
\newcommand{\TokRatioOneFiveOverZeroFive}{1.98}
\newcommand{\RuntimeGibMovedBudget}{1}
\newcommand{\NManifestsWord}{four}
\newcommand{\MarginalSolveRatioWord}{four}
\newcommand{\ThreeBQfiveExtraSolvesWord}{one}
\newcommand{\FrontierStableNWord}{five}
\newcommand{\NTasks}{542}
\newcommand{\NTasksHe}{164}
\newcommand{\NTasksMbpp}{378}
\newcommand{\NConfigs}{24}
\newcommand{\BootBJps}{100{,}000}
\newcommand{\BootSeed}{20260810}

\newcommand{\qtwo}{\texttt{q2\_K}\xspace}
\newcommand{\qthree}{\texttt{q3\_K\_M}\xspace}
\newcommand{\qfour}{\texttt{q4\_K\_M}\xspace}
\newcommand{\qfive}{\texttt{q5\_K\_M}\xspace}
\newcommand{\qsix}{\texttt{q6\_K}\xspace}
\newcommand{\qeight}{\texttt{q8\_0}\xspace}
\newcommand{\fp}{\texttt{fp16}\xspace}
\newcommand{\na}{$\cdot$}

\begin{document}
\ifdefined\pdfinfo\pdfinfo{/Title (Choosing Quantized Code LLMs by File Size and Energy per Solved Task) /Author (Aditi Patodiya)}\else\fi%

\title{Choosing Quantized Code LLMs by File Size and Energy per Solved Task%
\thanks{Submitted to GREENS 2027, the 11th International Workshop on Green and Sustainable Software at ICSE 2027. This work has been submitted to the IEEE for possible publication. Copyright may be transferred without notice, after which this version may no longer be accessible.}}

\author{\IEEEauthorblockN{Aditi Patodiya}\IEEEauthorblockA{Independent Researcher\\Email: aditi.patodia31@gmail.com}}

\maketitle

\begin{abstract}
A developer running a code assistant on a laptop downloads one model file, and local inference moves the energy of every request onto that machine.
A common rule says a larger model at lower precision tends to beat a smaller one at higher precision.
We ran \NConfigs{} GGUF files of Qwen2.5-Coder at four sizes on a fanless MacBook Air M2 and priced each in energy per solved task on \NTasks{} EvalPlus tasks.
The result is a selection guide indexed by Ollama's listed file size, with an evidence label on each pick.
At 1 and 2 GB it picks the \qfour files of the 1.5B and 3B models, both on the correctness and energy frontier under every accounting.
At 3 GB as listed, the larger-model rule picks 7B-\qtwo, the only measured 7B file that fits, and there it does not pay.
That file brings no significant gain in solves over the 2 GB pick, 3B-\qfour, and the 90\% CI limits any 3B-\qfour deficit to \HeadMaxDeficit{} points.
Yet 7B-\qtwo needs \HeadJpsRatio{} times the energy per solved task on CPU, GPU and memory rails, or \HeadJpsRatioSysLo{} to \HeadJpsRatioSysHi{} times on whole-system power.
On HumanEval+, five alternating session pairs give a ratio of \TtwoRRailsThree{} against \TtwoAugustRRails{} in the main grid.
At 8 GB, \BudgetEightPick{} buys a significant gain over the 2 GB pick at a higher energy price.
From 1.5B up, \qfour trails the highest precision measured at its size by less than \EquivMarginPP{} points (90\% CI).
We release per-task results, power logs and the analysis script.
\end{abstract}

\begin{IEEEkeywords}
green software, energy efficiency, large language models for code, quantization, GGUF k-quants, local inference, energy per solved task, model selection
\end{IEEEkeywords}

\section{Introduction}\label{sec:intro}

A developer who runs a code assistant on a laptop downloads one model file.
That file fixes a model size and a quantization recipe.
Ollama ships each size of a model family under several recipe tags~\cite{ollama}.
A tag names a recipe and not a bit count: the 0.5B-\qthree file stores \BpwZeroFiveBQthree{} effective bits per weight (BPW), and the 7B-\qtwo file stores \BpwSevenBQtwo{}.
The download is also an energy choice, because local inference moves the energy of code assistance onto the developer's machine.
Recent studies measure that energy for local code models and find that the choice of model moves it~\cite{cunha2026notall,alizadeh2025msr}.
Cunha et al.~\cite{cunha2026notall} guide the choice of model and prompt.
With the code family and prompt fixed, a developer still has to pick a file that fits a size budget, and that is the choice we price.

A common rule guides the size choice.
A larger model at lower precision tends to beat a smaller model at higher precision~\cite{song2025ondevice,saadfalcon2025ipw,alizadeh2025msr}, and at a fixed total of bits four-bit precision is almost universally optimal~\cite{dettmers2023kbit}.
Song et al.\ report a performance threshold for this rule near 3.5 effective BPW on general models~\cite{song2025ondevice}.
We ask where the rule fails to pay on these code tasks once energy per solved task is counted.
We price files in energy per solved task and treat it as the developer's cost, since a developer pays for working code.

We measured \NConfigs{} shipped GGUF files of Qwen2.5-Coder-Instruct at four sizes on a fanless MacBook Air M2.
Every file ran the same \NTasks{} EvalPlus tasks.
We compare solves task by task with exact paired tests.
For the key step of the guide, 3B-\qfour against 7B-\qtwo (the anchor pair), we also ran five alternating session pairs.

Table~\ref{tab:guide} gives the resulting guide.
At a 1 GB listed-size budget it picks \BudgetOnePick{}, and at 2 GB it picks \BudgetTwoPick{}.
Both stay on the correctness and energy frontier under every accounting.
At 3 GB as listed, the larger-model rule picks 7B-\qtwo, the only measured 7B file that fits, and there it does not pay.

\textbf{RQ1 (within a size).} What does each step down the quantization ladder cost in solved tasks and in energy per solved task at each of the four sizes?

\textbf{RQ2 (across sizes at a budget).} For a listed file-size budget, which file solves the most tasks, and which files sit on the correctness and energy frontier with a margin large enough to recommend them?

\textbf{RQ3 (the anchor step).} Where does the larger-model rule fail to pay, and does the energy price of that step hold across alternating repeated sessions and a change of energy accounting?

The paper makes three contributions.

\textbf{C1. A selection guide for local code LLMs.}
The guide is indexed by the file size Ollama lists, at budgets of 0.5, 1, 2, 4 and 8 GB.
For each budget it names the file that solves the most tasks among the \NConfigs{} tested, its energy per solved task and its in-budget runner-up.
Every budget pick and every rule that rests on a comparison carries an evidence label that says whether it comes from the repeat, a correctness test or a point-estimate energy margin (Table~\ref{tab:guide}, Sec.~\ref{sec:guide}).

\textbf{C2. The anchor step at 2 to 3 GB, priced and repeated.}
3B-\qfour solves \HeadSolvesA{} tasks against \HeadSolvesB{} for 7B-\qtwo, a file of \BpwSevenBQtwo{} effective BPW, below the threshold Song et al.\ report.
The 90\% CI limits any 3B-\qfour deficit to \HeadMaxDeficit{} points.
The 7B file needs \HeadJpsRatio{} times the energy per solved task (\HeadJpsRatioSysLo{} to \HeadJpsRatioSysHi{} under whole-system accounting).
On HumanEval+, five alternating session pairs give a ratio of \TtwoRRailsThree{} against \TtwoAugustRRails{} for those tasks in the main grid.

\textbf{C3. A regenerable price list for \NConfigs{} shipped files.}
For every file we report effective bits per weight and energy per solved task beside energy per token.
The released script regenerates every measured number.

\section{Background and Related Work}\label{sec:related}

\subsection{What a GGUF tag does and does not tell you}\label{sec:kquants}
The llama.cpp runtime runs quantized models stored as GGUF files~\cite{llamacpp}.
Its k-quant recipes mix several tensor types within one file~\cite{kawrakow2023kquants}.
At 1.5B and above, a tag such as \qfour names the recipe of most quantized tensors, while some stay at higher precision.
At 0.5B the tag names a minority: the \qfour file holds \TensorQfourZeroFiveB{} q4\_K tensors of \TensorQuantZeroFiveB{}, and the \qsix file holds \TensorQsixZeroFiveB{} q6\_K tensors, the rest q8\_0.
The effective BPW of a file therefore departs from the number in its tag.
The 0.5B-\qthree file holds no three-bit tensors and stores \BpwZeroFiveBQthree{} BPW.

\subsection{Energy and correctness of local code LLMs}\label{sec:prior}
Cunha et al.~\cite{cunha2026notall} benchmark five local LLMs on code tasks at GREENS 2026.
Each model runs as one GGUF file at its own quantization level, on HumanEval-X and MBPP+, with CPU-only inference through llama.cpp on an Intel Core i7-8700 machine.
They show that models trade accuracy against energy, and that the choice of model and of prompting both move energy and run time, so a developer should make the two choices together.
We hold one code family and one zero-shot prompt fixed and vary model size and k-quant recipe on the GPU of a fanless laptop.

Energy per solved task is already in use.
Abkin et al.~\cite{abkin2026energy} report energy per verified agent task with bootstrap intervals.
A TU Delft course study~\cite{tudelft2026g7} reports energy per correct MBPP+ solution, and Saad-Falcon et al.~\cite{saadfalcon2025ipw} report accuracy per joule.

Alizadeh et al.~\cite{alizadeh2025msr,alizadeh2026esem} relate energy to code accuracy for Ollama fp16, q8 and q4 tags on a laptop GPU and an A100, with Qwen2.5-Coder 1.5B and 7B in the later study.
They find quantized large models ahead of full-precision medium ones, and the later study finds most Pareto-optimal points quantized.
Our \fp result agrees on a fanless Apple laptop.
Their ladder stops at q4 and reports energy per token or per workload, while ours continues to \qthree and \qtwo and prices each file per solved task.
On Ollama, quantized tags used less energy than \fp~\cite{alizadeh2025msr,alizadeh2026esem}, while two studies on Hugging Face Transformers report that quantization raised energy~\cite{dereus2024starcoder2,vallecillos2026apr}.
Our files sit on the Ollama side.
Husom et al.~\cite{husom2025sustainable} profile quantized Ollama models at several quantization levels on a Raspberry Pi.
Rajput and Sharma find GGML and GGUF the most energy-efficient of five quantization methods~\cite{rajput2024quant}.
Wei et al.~\cite{wei2023greener} quantized code models to cut resource use and carbon footprint, and ran them on an M1 MacBook Pro.
Afrin et al.~\cite{afrin2026quantize} compare six quantization methods for Qwen2.5-Coder and CodeLlama.
Shi et al.\ (Ditto)~\cite{ditto2026} score EvalPlus tasks and estimate energy per token from throughput at a fixed 18~W on an Apple M4, where we read rail power per task.

Giagnorio et al.~\cite{giagnorio2025ptq} and Nyamsuren~\cite{nyamsuren2024lua} find four-bit precision a good trade for code and a steep loss at 2 bits.
Giagnorio et al.\ also find that large 2-bit models can beat smaller full-precision ones, with no general trend, and Nyamsuren finds small \fp models of similar size near the level of 2-bit 7B models on Lua.
Neither measures energy.
Thangarajah et al.\ serve Qwen2.5-Coder-3B through Ollama and find that \qfour keeps its \fp quality on code completion~\cite{synconfroute}.
Our ladder agrees at 3B.
At 7B it agrees with their advice for models of 3B or more.
At 1.5B, where they advise \qeight, we find \qfour equivalent to \fp within \EquivMarginPP{} points.
Melton~\cite{melton2026mlx} tests MLX four-bit files, including Qwen2.5-Coder, on the same \NTasks{} EvalPlus tasks without measuring energy, and finds a significant loss at 1.5B and 3B but not at 7B.
MLX also quantizes embeddings and the output head to four bits, while the \qfour files here store \BpwOneFiveBQfour{} and \BpwThreeBQfour{} effective bits per weight at those sizes, so two nominal four-bit files need not store the same bits.
Song et al.~\cite{song2025ondevice} find that heavily quantized larger models beat smaller high-precision ones on general models, with a performance threshold near 3.5 effective bits per weight.
Our 7B-\qtwo file stores \BpwSevenBQtwo{}, below that threshold.
Our question sits between these lines of work: one code family, priced in solved tasks and joules, file by file at a size budget.

\subsection{Energy measurement on macOS}\label{sec:macos}
Apple's powermetrics tool reports component power that its manual describes as estimated~\cite{powermetrics}.
The macmon tool reads the same IOReport counters without sudo~\cite{macmon}.
Cruz's guide asks energy experiments to repeat and shuffle their runs~\cite{cruz2021guide}.
In every repeat we ran, outputs repeated byte for byte (Sec.~\ref{sec:repeats}), so those repeats bought energy precision alone.
The paired repeat alternates the two files behind our selection guide's key rule.

\section{Study Design}\label{sec:design}

\subsection{Files}\label{sec:files}
We study Qwen2.5-Coder-Instruct at 0.5B, 1.5B, 3B and 7B parameters~\cite{hui2024qwen25coder}.
Every size runs \qthree, \qfour, \qfive, \qsix and \qeight.
The 0.5B to 3B sizes also run \fp, and 7B alone adds \qtwo, to trace the low end at the largest size.
That gives \NConfigs{} files.
7B was not run at \fp, which lists at about 15 GB on this 16 GB machine, so \qeight is its highest precision measured.
We use the registry tags as pulled and identify each file by its digest, because a GGUF file need not record how it was built~\cite{liutsai2026}.
Effective BPW comes from the runtime's load log, the listed size from \texttt{ollama list}, and file bytes from the registry manifests for \NManifestsWord{} files, the anchor pair among them, and from the load log plus metadata for the rest.
We index budgets by the size Ollama lists, because a developer sees it before download.
Budgets are file sizes, not memory, and Rule~\ref{rule:smoke} gives the headroom to leave for the runtime.

\subsection{Runtime and machine}\label{sec:runtime}
All files ran on Ollama 0.32.6 with its llama.cpp runner on Metal~\cite{ollama,llamacpp}.
The machine is a fanless MacBook Air M2 with 16 GB of unified memory, under macOS 26.5, on AC power.
Decoding takes the top-scoring token at temperature zero with a fixed seed, under Ollama's default repeat penalty of 1.1 over the last 64 tokens, and each task gets one sample.
We call this greedy decoding.
The context holds 2,048 tokens, and each answer has at most 768 new tokens.
Files ran one at a time in a fixed order, by size and then recipe, with one session per benchmark.
We call these runs the main grid.

\subsection{Tasks and scoring}\label{sec:tasks}
The tasks are EvalPlus HumanEval+ (\NTasksHe{} tasks) and MBPP+ (\NTasksMbpp{} tasks), \NTasks{} tasks pooled~\cite{liu2023evalplus,chen2021codex,austin2021mbpp}.
A task is solved when its sanitized solution passes both the base and the plus tests of EvalPlus.
Correctness comes from a serial re-evaluation of the stored solutions, covering every failed task and a seeded 3\% sample of passes, not new generation.
It changed \SerialChangedOutcomes{} outcomes in \SerialChangedConfigs{} files, all from fail to pass, with \SerialHeChanges{} on HumanEval+.
No sampled pass flipped, which supports leaving the other passes as scored.
These correctness observations also supplied the tuning data of our registry-screening study~\cite{patodiya2026p2}.
This paper adds the energy and file-selection analysis of those runs and the paired energy repeat.

\subsection{Energy}\label{sec:energy}
In the main grid the macmon tool samples the CPU, GPU and memory rails once per second~\cite{macmon}.
A task window runs from the request to the response.
We use a count rule: each macmon sample inside a task window counts once, at 1 s.
Marginal power is rail power above an idle median of \IdleRailsWThree{} W.
Energy per solved task (J/solve) is the pooled marginal energy of both benchmark runs divided by the pooled solves.
We report energy per output token (mJ/token) beside it.
J/solve intervals are 95\% task-bootstrap intervals within the main-grid sessions (B~=~\BootBJps{}, seed \BootSeed{}), printed in whole joules.
For a whole-system view we use macmon's system power under three idle choices: a pooled median, a clean-idle median and no subtraction.
In that view each sample is weighted by its gap to the previous one, with long gaps capped.
We print that view as a range, and we say \emph{under every accounting} when a point estimate holds across the rails and this view, under each idle choice and sample weighting we computed.
Gap weighting moves J/solve by under \GapWeightBoundPct\% on every file and leaves the rails frontier unchanged.

\subsection{Statistics and the headline rule}\label{sec:stats}
Paired tests of no difference use the exact McNemar test on the discordant tasks~\cite{mcnemar1947}.
Holm correction~\cite{holm1979} runs within each size for adjacent steps.
Over the four sizes it runs in two separate families: the endpoints, and \qfour against the highest precision measured.
Budget picks and the gains against the 2 GB pick are post hoc and chosen from the same data, so we correct them with Holm over all \AllPairsN{} pairs of the \NConfigs{} files, which covers any pair chosen after seeing the results.
Intervals on pass-rate differences come from a task bootstrap computed exactly, with no Monte Carlo error, and are unadjusted.
Equivalence (two one-sided tests~\cite{schuirmann1987}) and non-inferiority use a margin of \EquivMarginPP{} points (about \EquivMarginTasks{} of \NTasks{} tasks), chosen after seeing the data.
It lies below the smallest gain from 3B to 7B at a matched recipe (\SizeGainThreeSevenMin{} points), and every verdict in Table~\ref{tab:tests} is unchanged for any margin from \EquivMarginStableLo{} to \EquivMarginStableHi{} points.
Two files are equivalent when the 90\% CI of their difference, which matches the two one-sided 5\% tests, lies inside $\pm$\EquivMarginPP{} points.
\qfour is non-inferior when the upper 90\% bound of the top-minus-\qfour difference is below \EquivMarginPP{} points.
These readings rest on unadjusted intervals.
Over the four sizes under Holm, a paired score test~\cite{tango1998} of non-inferiority gives adjusted p \NonInfScoreHolmOneFiveB{}, \NonInfScoreHolmThreeB{} and \NonInfScoreHolmSevenB{} at 1.5B, 3B and 7B.
The bootstrap tail at +\EquivMarginPP{} points gives \NonInfPHolmOneFiveB{}, \NonInfPHolmThreeB{} and \NonInfPHolmSevenB{}.
Simultaneous evidence at 1.5B and 7B therefore depends on the method.
For the anchor pair, 3B-\qfour is non-inferior when the lower 90\% bound of 3B minus 7B lies above $-$\EquivMarginPP{} points.
Significance rests on the Holm-adjusted p values.

The frontier is the non-dominated set over all \NConfigs{} files on pooled pass rate (higher is better) and J/solve (lower is better).
A frontier file's margin is how far its J/solve could rise before another file dominates it.
An off-frontier file's margin is how far its J/solve would have to fall to join.
A step to a larger model pays when no smaller file dominates the file it buys.

We headline an energy difference only if its margin is at least 40\% or the paired repeat covers it.
This is an exploratory screening heuristic that we set after the main grid.
Speed can move between sessions with outputs fixed: a repeat of 1.5B-\qfour on HumanEval+, run without power logging, reproduced \RepeatIdentical{} of \NTasksHe{} outputs byte for byte, while median decode speed moved \RepeatDecodePct\% and wall time \RepeatWallPct\%.
As a rough calibration, each file's two power-logged sessions, which also differ in benchmark, differ in rail energy per token by at most \SessionTokSwing\% (median \SessionTokSwingMedian\%).
Compounded over both files of a comparison, that swing is a factor of \SessionTokSwingFactor{}, and every margin we headline clears it as a ratio (smallest \RuleMarginRatioMin{}).
The largest block swing in the paired repeat, \TtwoBlockSwingMax{} under macOS background work, stays below that factor.
The labels do not change for any threshold from \RuleStableLo\% to \RuleStableHi\%.

\subsection{Repeats}\label{sec:repeats}
In every repeat we ran, the outputs were byte-identical under greedy decoding.
That holds for \RepeatIdentical{} of \NTasksHe{} outputs in the 1.5B repeat, for all \TtwoValidBlocks{} blocks of the paired repeat against the main grid, and for all \BrokenRepeatIdentical{} outputs of a 3B-\qthree repeat on HumanEval+.
The guide's picks are the files that solve the most tasks, so in these repeats session effects touched energy and speed, not the picks.

The grid's fixed order ties size to time of day, so the paired repeat alternates the two anchor files.
We chose this pair after the main grid: it joins the 2 GB pick to the smallest 7B file, the first file of the larger model above that budget.
The repeat ran five alternating pairs (AB BA AB BA AB) of 3B-\qfour (A) and 7B-\qtwo (B) in one night.
Each letter is one block, a run of one file over all \NTasksHe{} tasks, preceded by a cooldown gate of at least five minutes, until the CPU reached 45$^\circ$C or below.
The repeat runs on HumanEval+, the benchmark with the lower main-grid ratio (\HeadJpsRatioHe{} against \HeadJpsRatioMbpp{} on MBPP+ under the count rule), so it tests the smaller gap.
Repeat energy, sampled about every 200 ms, uses the gap weighting of Sec.~\ref{sec:energy}, and so does the main-grid reference value for the repeat.
A block was valid if it ran all \NTasksHe{} tasks with outputs byte-identical to the main grid, stayed on AC power throughout, and had power samples over at least \TtwoValidityCoveragePct\% of the duration of its task windows.
A valid block was flagged if it was charging, the display was on, thermal pressure rose above zero or Low Power Mode was on.
The validity rule, the flag rule and the analysis were fixed before the run, in the released scripts.
The analysis is the geometric mean of the per-pair J/solve ratio R (B over A), with a 95\% t-interval on log R.
All \TtwoValidBlocks{} blocks were valid.

\section{Results}\label{sec:results}

\begin{table*}[t]
\caption{The \NConfigs{} files: effective bits per weight, listed size, pooled pass rate, energy per solved task and energy per output token}
\label{tab:files}
\centering
\footnotesize
\setlength{\tabcolsep}{4.5pt}
\begin{tabular}{@{}lrrrrrrrrrrrrrrrrrrrr@{}}
\toprule
 & \multicolumn{5}{c}{0.5B} & \multicolumn{5}{c}{1.5B} & \multicolumn{5}{c}{3B} & \multicolumn{5}{c}{7B} \\
\cmidrule(lr){2-6}\cmidrule(lr){7-11}\cmidrule(lr){12-16}\cmidrule(l){17-21}
Recipe & BPW & GB & Pass & J & mJ & BPW & GB & Pass & J & mJ & BPW & GB & Pass & J & mJ & BPW & GB & Pass & J & mJ \\
\midrule
\qtwo & \na & \na & \na & \na & \na & \na & \na & \na & \na & \na & \na & \na & \na & \na & \na & \BpwSevenBQtwo & \ListedGBSevenBQtwo & \PassSevenBQtwo & \JpsSevenBQtwo & \MjTokSevenBQtwo \\
\qthree & \BpwZeroFiveBQthree & \ListedGBZeroFiveBQthree & \PassZeroFiveBQthree & \JpsZeroFiveBQthree & \MjTokZeroFiveBQthree & \BpwOneFiveBQthree & \ListedGBOneFiveBQthree & \PassOneFiveBQthree & \JpsOneFiveBQthree & \MjTokOneFiveBQthree & \BpwThreeBQthree & \ListedGBThreeBQthree & \PassThreeBQthree & \JpsThreeBQthree & \MjTokThreeBQthree & \BpwSevenBQthree & \ListedGBSevenBQthree & \PassSevenBQthree & \JpsSevenBQthree & \MjTokSevenBQthree \\
\qfour & \BpwZeroFiveBQfour & \ListedGBZeroFiveBQfour & \PassZeroFiveBQfour & \JpsZeroFiveBQfour & \MjTokZeroFiveBQfour & \BpwOneFiveBQfour & \ListedGBOneFiveBQfour & \PassOneFiveBQfour & \JpsOneFiveBQfour & \MjTokOneFiveBQfour & \BpwThreeBQfour & \ListedGBThreeBQfour & \PassThreeBQfour & \JpsThreeBQfour & \MjTokThreeBQfour & \BpwSevenBQfour & \ListedGBSevenBQfour & \PassSevenBQfour & \JpsSevenBQfour & \MjTokSevenBQfour \\
\qfive & \BpwZeroFiveBQfive & \ListedGBZeroFiveBQfive & \PassZeroFiveBQfive & \JpsZeroFiveBQfive & \MjTokZeroFiveBQfive & \BpwOneFiveBQfive & \ListedGBOneFiveBQfive & \PassOneFiveBQfive & \JpsOneFiveBQfive & \MjTokOneFiveBQfive & \BpwThreeBQfive & \ListedGBThreeBQfive & \PassThreeBQfive & \JpsThreeBQfive & \MjTokThreeBQfive & \BpwSevenBQfive & \ListedGBSevenBQfive & \PassSevenBQfive & \JpsSevenBQfive & \MjTokSevenBQfive \\
\qsix & \BpwZeroFiveBQsix & \ListedGBZeroFiveBQsix & \PassZeroFiveBQsix & \JpsZeroFiveBQsix & \MjTokZeroFiveBQsix & \BpwOneFiveBQsix & \ListedGBOneFiveBQsix & \PassOneFiveBQsix & \JpsOneFiveBQsix & \MjTokOneFiveBQsix & \BpwThreeBQsix & \ListedGBThreeBQsix & \PassThreeBQsix & \JpsThreeBQsix & \MjTokThreeBQsix & \BpwSevenBQsix & \ListedGBSevenBQsix & \PassSevenBQsix & \JpsSevenBQsix & \MjTokSevenBQsix \\
\qeight & \BpwZeroFiveBQeight & \ListedGBZeroFiveBQeight & \PassZeroFiveBQeight & \JpsZeroFiveBQeight & \MjTokZeroFiveBQeight & \BpwOneFiveBQeight & \ListedGBOneFiveBQeight & \PassOneFiveBQeight & \JpsOneFiveBQeight & \MjTokOneFiveBQeight & \BpwThreeBQeight & \ListedGBThreeBQeight & \PassThreeBQeight & \JpsThreeBQeight & \MjTokThreeBQeight & \BpwSevenBQeight & \ListedGBSevenBQeight & \PassSevenBQeight & \JpsSevenBQeight & \MjTokSevenBQeight \\
\fp & \BpwZeroFiveBFpsixteen & \ListedGBZeroFiveBFpsixteen & \PassZeroFiveBFpsixteen & \JpsZeroFiveBFpsixteen & \MjTokZeroFiveBFpsixteen & \BpwOneFiveBFpsixteen & \ListedGBOneFiveBFpsixteen & \PassOneFiveBFpsixteen & \JpsOneFiveBFpsixteen & \MjTokOneFiveBFpsixteen & \BpwThreeBFpsixteen & \ListedGBThreeBFpsixteen & \PassThreeBFpsixteen & \JpsThreeBFpsixteen & \MjTokThreeBFpsixteen & \na & \na & \na & \na & \na \\
\bottomrule
\end{tabular}

\vspace{2pt}
\parbox[t]{\textwidth}{\footnotesize BPW: effective bits per weight from the load log. GB: size listed by Ollama. Pass: pooled pass rate over \NTasks{} tasks (\%). J: rail energy per solved task (J). mJ: rail energy per output token (mJ). One main-grid session per benchmark. A dot marks a recipe that was not run.}
\end{table*}

\begin{table*}[t]
\caption{The selection guide: the measured file that solves the most tasks within each listed budget}
\label{tab:guide}
\centering
\footnotesize
\setlength{\tabcolsep}{2.5pt}
\begin{tabular}{@{}l >{\raggedright\arraybackslash}p{2.6cm} l r >{\raggedright\arraybackslash}p{5.55cm} >{\raggedright\arraybackslash}p{5.95cm}@{}}
\toprule
Budget & Pick (listed GB) & Solved (pass) & J/solve & In-budget comparison (solved, J/solve) & Evidence label \\
\midrule
0.5 GB & \BudgetHalfPick{} (\ListedGBZeroFiveBQfive{}) & \SolvesZeroFiveBQfive{} (\PassZeroFiveBQfive\%) & \JpsZeroFiveBQfive{} & 0.5B-\qfour: \SolvesZeroFiveBQfour{}, \JpsZeroFiveBQfour{}~J; 0.5B-\qthree: \SolvesZeroFiveBQthree{}, \JpsZeroFiveBQthree{}~J & lead unresolved (raw $p$ \PickPHalf{}) \\
\addlinespace
1 GB & \BudgetOnePick{} (\ListedGBOneFiveBQfour{}) & \SolvesOneFiveBQfour{} (\PassOneFiveBQfour\%) & \JpsOneFiveBQfour{} & 1.5B-\qthree: \SolvesOneFiveBQthree{}, \JpsOneFiveBQthree{}~J (off the frontier by \FrontierMarginOneFiveBQthree\%) & lead unresolved (raw $p$ \PickPOne{}); frontier member, rails frontier margin \FrontierMarginOneFiveBQfour\% against 3B-\qfour \\
\addlinespace
2 GB & \BudgetTwoPick{} (\ListedGBThreeBQfour{}) & \SolvesThreeBQfour{} (\PassThreeBQfour\%) & \JpsThreeBQfour{} & 1.5B-\qfour: \SolvesOneFiveBQfour{}, \JpsOneFiveBQfour{}~J (\IsoThreeBOverOneFive{} points fewer). Trap: 3B-\qthree{} (\ListedGBThreeBQthree{}~GB) solves \SolvesThreeBQthree{}~\cite{patodiya2026p2} & significant lead (all-pairs Holm $p$ \AllPairsPTwo{}); frontier member, rails frontier margin \FrontierMarginThreeBQfour\% against 3B-\qfive \\
\addlinespace
4 GB & \BudgetFourPick{} (\ListedGBSevenBQthree{}) & \SolvesSevenBQthree{} (\PassSevenBQthree\%) & \JpsSevenBQthree{} & 3B-\qfive: \SolvesThreeBQfive{}, \JpsThreeBQfive{}~J; 3B-\qfour: \SolvesThreeBQfour{}, \JpsThreeBQfour{}~J; 7B-\qtwo: \SolvesSevenBQtwo{}, \JpsSevenBQtwo{}~J & lead unresolved (raw $p$ \PickPFour{}); off the frontier by \FrontierMarginSevenBQthree\%, unreplicated \\
\addlinespace
8 GB & \BudgetEightPick{} (\ListedGBSevenBQsix{}) & \SolvesSevenBQsix{} (\PassSevenBQsix\%) & \JpsSevenBQsix{} & 7B-\qfour{} (\ListedGBSevenBQfour{}): \SolvesSevenBQfour{}, \JpsSevenBQfour{}~J, smallest file shown equivalent to \qeight{} (\ListedGBSevenBQeight{}, over budget) & lead unresolved from 7B-\qfour to \qsix (raw $p$ \PickPEight{}) \\
\bottomrule
\end{tabular}

\vspace{2pt}
\parbox[t]{\textwidth}{\footnotesize Listed size is file size, not memory. Every row comes from single sessions; only the anchor pair behind Rule~\ref{rule:anchor} is repeated. A row's $p$ is a post hoc exact McNemar test of the pick against its in-budget runner-up, the first file in its comparison cell. \emph{Repeated} means the paired repeat covers the rule. \emph{Margin $\geq$40\%} means a frontier margin or energy gap of 40\% or more on the rails, from single sessions, an exploratory screen. A \emph{frontier member} stays on the point-estimate frontier under every accounting. A \emph{significant lead} is a solve difference that stays significant after Holm over all \AllPairsN{} file pairs. Any other lead is \emph{unresolved}, which does not establish equivalence. An energy ranking is \emph{unreplicated} when the gap is below 40\% and the repeat does not cover it.}
\end{table*}

\begin{table}[t]
\caption{Paired tests and energy ratios behind the guide}
\label{tab:tests}
\centering
\footnotesize
\setlength{\tabcolsep}{2.5pt}
\begin{tabular}{@{}lrlr>{\raggedright\arraybackslash}p{2.2cm}@{}}
\toprule
Comparison & Diff. & Interval & $p$ & Verdict \\
\midrule
\multicolumn{5}{@{}l}{\emph{A. Top precision minus \qfour (90\% CI; post hoc margin)}} \\
0.5B, \fp & \QfourTopDiffZeroFiveB{} & [\QfourTopCINinetyLoZeroFiveB{}, \QfourTopCINinetyHiZeroFiveB{}] & \QfourTopPHolmZeroFiveB{} & \QfourTopVerdictZeroFiveB{} \\
1.5B, \fp & \QfourTopDiffOneFiveB{} & [\QfourTopCINinetyLoOneFiveB{}, \QfourTopCINinetyHiOneFiveB{}] & \QfourTopPHolmOneFiveB{} & \QfourTopVerdictOneFiveB{} \\
3B, \fp & \QfourTopDiffThreeB{} & [\QfourTopCINinetyLoThreeB{}, \QfourTopCINinetyHiThreeB{}] & \QfourTopPHolmThreeB{} & \QfourTopVerdictThreeB{} \\
7B, \qeight & \QfourTopDiffSevenB{} & [\QfourTopCINinetyLoSevenB{}, \QfourTopCINinetyHiSevenB{}] & \QfourTopPHolmSevenB{} & \QfourTopVerdictSevenB{} \\
\midrule
\multicolumn{5}{@{}l}{\emph{B. Top precision minus lowest rung (95\% CI)}} \\
0.5B, \qthree & \EndDiffZeroFiveB{} & [\EndCILoZeroFiveB{}, \EndCIHiZeroFiveB{}] & \EndPHolmZeroFiveB{} & \EndSigHolmZeroFiveB{} \\
1.5B, \qthree & \EndDiffOneFiveB{} & [\EndCILoOneFiveB{}, \EndCIHiOneFiveB{}] & \EndPHolmOneFiveB{} & \EndSigHolmOneFiveB{} \\
3B, \qthree{}$^{a}$ & \EndDiffThreeB{} & [\EndCILoThreeB{}, \EndCIHiThreeB{}] & \EndPHolmThreeB{} & \EndSigHolmThreeB{} \\
7B, \qtwo & \EndDiffSevenB{} & [\EndCILoSevenB{}, \EndCIHiSevenB{}] & \EndPHolmSevenB{} & \EndSigHolmSevenB{} \\
\midrule
\multicolumn{5}{@{}l}{\emph{C. Anchor pair, 3B-\qfour against 7B-\qtwo}} \\
Solved, 3B / 7B & \multicolumn{4}{l}{\HeadSolvesA{} / \HeadSolvesB{}} \\
Overlap$^{b}$ & \multicolumn{4}{l}{\HeadOverlapBoth{} / \HeadOverlapOnlyA{} / \HeadOverlapOnlyB{} / \HeadOverlapNeither{}} \\
3B minus 7B & \HeadDiffPP{} & [\HeadCINinetyLo{}, \HeadCINinetyHi{}]$^{c}$ & \HeadP{}$^{d}$ & non-inferior$^{e}$ \\
J/solve ratio & \HeadJpsRatio{} & [\HeadJpsRatioCILo{}, \HeadJpsRatioCIHi{}]$^{f}$ & & \\
System power & \multicolumn{2}{l}{\HeadJpsRatioSysLo{} to \HeadJpsRatioSysHi{}$^{g}$} & & \\
HE+ repeat & \TtwoRRailsThree{} & [\TtwoRRailsLoThree{}, \TtwoRRailsHiThree{}]$^{h}$ & & repeated \\
HE+ main grid$^{i}$ & \TtwoAugustRRails{} & & & \\
\midrule
\multicolumn{5}{@{}l}{\emph{D. 7B, \qthree minus \qtwo (95\% CI), for context}} \\
\qtwo to \qthree & \AdjDiffSevenBQtwoQthree{} & [\AdjCILoSevenBQtwoQthree{}, \AdjCIHiSevenBQtwoQthree{}] & \AdjPHolmSevenBQtwoQthree{} & \AdjSigHolmSevenBQtwoQthree{} \\
\midrule
\multicolumn{5}{@{}l}{\emph{E. \fp off the frontier (rails margin; system range)}} \\
1.5B, \fp & \MarginFpOffOneFiveBCount\% & \MarginFpOffOneFiveBSysLo{} to \MarginFpOffOneFiveBSysHi\%$^{g}$ & & margin $\geq$40\% \\
3B, \fp & \MarginFpOffThreeBCount\% & \MarginFpOffThreeBSysLo{} to \MarginFpOffThreeBSysHi\%$^{g}$ & & margin $\geq$40\% \\
\bottomrule
\end{tabular}

\vspace{2pt}
\parbox[t]{\columnwidth}{\footnotesize Diff.\ in points, except the energy ratios in C (7B over 3B) and the rails margins in E. Top precision is \fp, and \qeight at 7B. In A, B and D, $p$ is the Holm-adjusted exact McNemar p, intervals are unadjusted task-bootstrap intervals, and the B and D verdict says whether the step is significant after Holm. In A that $p$ tests the difference, and the verdict comes from the interval and the margin. HE+: HumanEval+. $^{a}$See~\cite{patodiya2026p2} for this file. $^{b}$Solved by both, by 3B only, by 7B only and by neither. $^{c}$90\% CI. $^{d}$Exact McNemar p for the single pair. $^{e}$At the post hoc margin; equivalence not shown because the upper bound favors 3B. $^{f}$95\% within-session task bootstrap over \NTasks{} tasks. $^{g}$System power, range over three idle choices. $^{h}$95\% between-session interval, five pairs. $^{i}$Gap-weighted, as in the repeat.}
\end{table}

\subsection{RQ1: \qfour holds from 1.5B up}\label{sec:rq1}
From 1.5B up, \qfour trails the highest precision measured at its size by less than \EquivMarginPP{} points (90\% CI; Table~\ref{tab:tests}, part A).
Within that margin \qfour is equivalent to \fp at 1.5B and to \qeight, the highest precision measured, at 7B.
At 3B it is non-inferior to \fp: it trails by at most \QfourTopQfourWorseUpToThreeB{} points and may be up to \QfourTopQfourBetterUpToThreeB{} points better.
At 0.5B, non-inferiority is not shown, and the loss is not significant after Holm (p \QfourTopPHolmZeroFiveB{}; raw p \QfourTopPZeroFiveB{}).

Part B of Table~\ref{tab:tests} asks what the whole ladder costs, from the highest precision down to the lowest rung, and the answer differs by size.
The difference is significant after Holm at 0.5B (\EndDiffZeroFiveB{} points, \fp against \qthree) and at 7B (\EndDiffSevenB{} points, \qeight against \qtwo), and not at 1.5B (\EndDiffOneFiveB{} points).
At 7B, the step from \qtwo to \qthree (raw p \AdjPSevenBQtwoQthree{}) is not significant after Holm (part D).
The 3B-\qthree file solves \SolvesThreeBQthree{} of \NTasks{} tasks, so the 3B difference, \fp against \qthree, is \EndDiffThreeB{} points.
For this model the registry file is tensor-identical to a plain uncalibrated llama.cpp rebuild, and importance-matrix quantization partly restores it~\cite{patodiya2026p2}.
The file stays in every statistic.

Within each size, the cheapest file per solve leads the next one by \CheapestGapZeroFiveB\% at 0.5B, \CheapestGapOneFiveB\% at 1.5B, \CheapestGapThreeB\% at 3B and \CheapestGapSevenB\% at 7B.
All four gaps fall below 40\%, so this stands as a pattern: \qfour is among the cheapest working files per solve at 1.5B and 3B, and at 7B it costs \GapSevenBQfourOverQtwo\% more per solve than \qtwo.

\begin{figure}[t]
\centering
\includegraphics[width=\columnwidth]{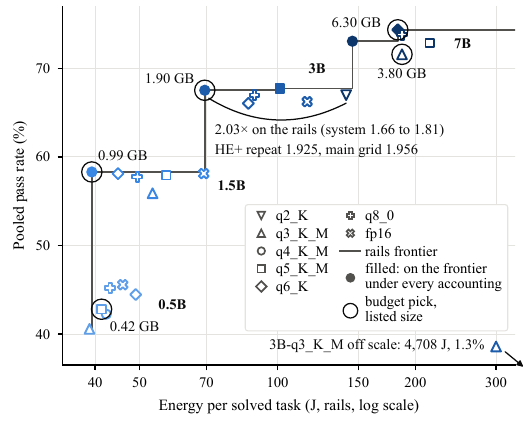}
\caption{Pooled pass rate against energy per solved task (log scale) for all \NConfigs{} files, each from its main-grid sessions. The step line is the rails frontier, and filled markers stay on it under every accounting. Rings mark the five budget picks with their listed size. A labeled link joins the anchor pair, 3B-\qfour and 7B-\qtwo. The 3B-\qthree file sits off scale (\JpsBrokenThreeB{}~J, \PassThreeBQthree\%).}
\label{fig:frontier}
\end{figure}

\subsection{RQ2: The 1 GB and 2 GB picks hold the frontier}\label{sec:rq2}
Within each listed budget the file that solves the most tasks is \BudgetHalfPick{}, \BudgetOnePick{}, \BudgetTwoPick{}, \BudgetFourPick{} and \BudgetEightPick{} (Table~\ref{tab:guide}).
The 1 GB and 2 GB picks also stay on the point-estimate frontier under every accounting (Table~\ref{tab:files}, Fig.~\ref{fig:frontier}).
The rails frontier has \FrontierN{} members: \FrontierList{}.
Under system power the low end changes: 0.5B-\qthree leaves and the frontier has \FrontierNSys{} members.
The other \FrontierStableNWord{} rails-frontier members stay on it under every accounting.
7B-\qsix is on the frontier only because it has the top pass rate.

At 1.5B and 3B, \fp stays off the frontier under every accounting.
On the rails it would need \MarginFpOffOneFiveBCount\% and \MarginFpOffThreeBCount\% less energy per solve to join.
Table~\ref{tab:tests}, part E, gives the system-power margins, and at 1.5B the dominating file there is \qsix.
7B-\qtwo sits off the frontier: 3B-\qfour dominates it, and the 7B file's energy per solve would have to fall \FrontierMarginSevenBQtwo\% to join.
Three other comparisons are too close to rank without repeats (Rule~\ref{rule:ties}).

At matched recipes from \qfour to \qeight, the 7B file costs \ScaleRatioMin{} to \ScaleRatioMax{} times the energy per solve of its 3B sibling for \SizeGainThreeSevenMin{} to \SizeGainThreeSevenMax{} more points, and at \qthree the ratio is \ScaleRatioQthree{}, because the 3B-\qthree file solves so few tasks.
3B-\qfour reaches \BestValueScorePct\% of the best score measured, that of \BestValueTop{}, at \BestValueEnergyPct\% of that file's energy per solve.

Computed in file bytes (decimal GB), the picks at all five budgets are unchanged, and Rule~\ref{rule:smoke} gives the three that move once the runtime's logged buffers count.
The \FrontierStableNWord{} stable members stay on the frontier when energy per task or per output token replaces energy per solve, and only the 0.5B end changes.
Over working files the ranks of energy per solve stay close to those of energy per token, and the difference falls where the budgets are.
Per token, 1.5B-\qfour costs \TokRatioOneFiveOverZeroFive{} times 0.5B-\qthree, while per solve the two sit within \MarginZeroFiveBQthreeOn\%, and \TokFrontierHalfBExtraWord{} 0.5B files that a per-token frontier would keep fall off the per-solve frontier.

\begin{figure}[t]
\centering
\includegraphics[width=\columnwidth]{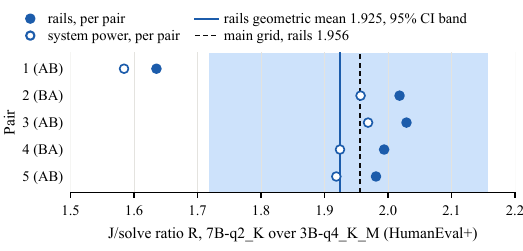}
\caption{Paired repeat on HumanEval+: per-pair J/solve ratio R of 7B-\qtwo over 3B-\qfour on the rails for five alternating pairs (AB or BA marked), with the pre-specified geometric mean and its 95\% interval. The reference line is the main-grid HumanEval+ rails value \TtwoAugustRRails{}. Hollow markers show the repeat under system power, with the display off. Pair 1 overlapped macOS background work, and the post hoc value without it is \TtwoRRailsNoPairOne{}.}
\label{fig:tier2}
\end{figure}

\subsection{RQ3: At 3 GB the larger model does not pay}\label{sec:rq3}
At 3 GB as listed, the larger-model rule picks 7B-\qtwo, the only measured 7B file that fits, and there it does not pay (Table~\ref{tab:tests}, part C).
The rule pays below this step: 3B-\qfour solves \SolvesThreeBQfour{} tasks against \SolvesOneFiveBQfour{} for 1.5B-\qfour (post hoc, all-pairs Holm p \AllPairsPTwo{}).
Against 7B-\qtwo, 3B-\qfour solves \HeadSolvesHeA{} to \HeadSolvesHeB{} on HumanEval+, and both solve \HeadSolvesMbppA{} on MBPP+.
As in Vallecillos-Ruiz et al.~\cite{vallecillos2026apr}, we report solved-set overlap: \HeadOverlapOnlyA{} tasks are solved only by 3B-\qfour and \HeadOverlapOnlyB{} only by 7B-\qtwo (exact McNemar p \HeadP{}).
The pass-rate difference, 3B minus 7B, is \HeadDiffPP{} points (90\% CI \HeadCINinetyLo{} to \HeadCINinetyHi{}; 95\% CI \HeadCILo{} to \HeadCIHi{}).
3B-\qfour is non-inferior at the post hoc \EquivMarginPP{}-point margin (paired score test~\cite{tango1998}, unadjusted one-sided p \AnchorNonInfScoreP{}).
Equivalence is not shown, but only because the interval's upper end, \HeadCINinetyHi{}, runs past \EquivMarginPP{} points in the 3B file's favor.
The extra file size thus buys no significant gain in solves.
3B-\qfour needs \HeadJpsA{}~J per solved task (95\% CI \JpsCILoThreeBQfour{} to \JpsCIHiThreeBQfour{}) and 7B-\qtwo needs \HeadJpsB{}~J (\JpsCILoSevenBQtwo{} to \JpsCIHiSevenBQtwo{}).
The ratio is \HeadJpsRatio{} (within-session task-bootstrap 95\% CI \HeadJpsRatioCILo{} to \HeadJpsRatioCIHi{}).
Under whole-system accounting it is \HeadJpsRatioSysLo{} to \HeadJpsRatioSysHi{}.
With near-equal solve counts, the per-solve gap follows the per-token cost of the larger file (Table~\ref{tab:files}).

The 7B-\qtwo file stores \BpwSevenBQtwo{} effective bits per weight, below the threshold near 3.5 that Song et al.\ report for general models~\cite{song2025ondevice}.
Within 7B, the step up to \qthree at \BpwSevenBQthree{} BPW is \AdjDiffSevenBQtwoQthree{} points, which is not significant after Holm (p \AdjPHolmSevenBQtwoQthree{}).
In Song et al.'s general Qwen 2.5 table, the counterpart of the larger anchor file still leads on four of five benchmarks, including HumanEval (53.66 to 48.78).
In the coder family on these tasks, 3B-\qfour solves \HeadSolvesA{} tasks to \HeadSolvesB{} (exact McNemar p \HeadP{}), so no significant lead remains and the developer still pays the energy.

As listed, the 3B-\qfour file is \ListedGBThreeBQfour{}~GB and the 7B-\qtwo file \ListedGBSevenBQtwo{}~GB, so the 3B file is \HeadListedLessPct\% smaller and the 7B file \HeadListedMorePct\% larger.
In file bytes they are \FileBytesGBThreeBQfour{} and \FileBytesGBSevenBQtwo{}~GB.
Near the same file size, 7B-\qtwo and 3B-\qeight (\ListedGBThreeBQeight{}~GB listed, \BpwThreeBQeight{} BPW) solve \SolvesSevenBQtwo{} tasks each, at \JpsSevenBQtwo{} and \JpsThreeBQeight{}~J per solve in single sessions.
Their solved sets split \NearPairSevenOnly{} to \NearPairThreeOnly{} (exact McNemar p \NearPairP{}).

On HumanEval+, the energy price of the step holds across alternating sessions (Fig.~\ref{fig:tier2}).
7B-\qtwo cost more per solved task in all five pairs.
The five pairs give a ratio of \TtwoRRailsThree{} (95\% CI \TtwoRRailsLoThree{} to \TtwoRRailsHiThree{}), against \TtwoAugustRRails{} for the main-grid HumanEval+ sessions with the same gap weighting.
The five-pair mean lies \TtwoBelowMainPct\% below that value, and a post hoc value without pair 1 (Sec.~\ref{sec:threats}) lies \TtwoAboveMainPct\% above it.
Without pair 1, the AB and BA pairs give similar means, \TtwoNoPairOneAB{} and \TtwoNoPairOneBA{}.
Under system power the paired-repeat ratio is \TtwoRSys{} (95\% CI \TtwoRSysLo{} to \TtwoRSysHi{}).
In the paired-repeat sessions 3B-\qfour decoded \TtwoDecode{} times as fast as 7B-\qtwo (95\% CI \TtwoDecodeLo{} to \TtwoDecodeHi{}), and 7B-\qtwo took \TtwoWall{} times the wall time (\TtwoWallLo{} to \TtwoWallHi{}).
Energy, decode speed and wall time moved by about the same factor, so in these sessions the larger file costs more because it runs longer at a similar rail power.
The cheaper file was also the faster one, so picking it gave up no response time.

\section{The Selection Guide}\label{sec:guide}
Four prices decide the download.
From the 1 GB pick to the 2 GB pick, \JpsRatioTwoOverOne{} times the energy per solve buys \IsoThreeBOverOneFive{} more points.
From the 2 GB pick to 7B-\qtwo, \HeadJpsRatio{} times (\HeadJpsRatioSysLo{} to \HeadJpsRatioSysHi{} under whole-system accounting) buys no significant gain.
From the 2 GB pick, 7B-\qfour costs \ScaleRatioQfour{} times for \ScaleGainFour{} more points and \BudgetEightPick{} costs \ScaleRatioSixFour{} times for \ScaleGainSixFour{} more.
Table~\ref{tab:guide} gives the picks.

A developer who budgets joules as well as gigabytes has a reason to stop at 3B-\qfour.
The next file up in solves, 3B-\qfive, costs \FrontierMarginThreeBQfour\% more per solve for \ThreeBQfiveExtraSolvesWord{} more task.
Along the stable frontier, in single sessions, each extra solved task costs about \MarginalSolveA{}~J from 1.5B-\qfour to 3B-\qfour and about \MarginalSolveB{}~J from 3B-\qfour to 7B-\qfour, about \MarginalSolveRatioWord{} times as much, and more under whole-system accounting.

\begin{enumerate}
\item\label{rule:tag} \emph{Read the listed size and the measured result, not the tag.}
The 0.5B-\qthree file stores \BpwZeroFiveBQthree{} effective bits per weight and the 7B-\qtwo file \BpwSevenBQtwo{}, and Rule~\ref{rule:smoke} shows a tag whose file fails most tasks, so the digits in a tag give neither the bits a file stores nor its correctness.

\item\label{rule:one} \emph{At a 1 GB budget, take \BudgetOnePick{}} [lead unresolved; frontier member].
It stays on the frontier under every accounting, and Table~\ref{tab:guide} gives its frontier margin and in-budget runner-up.

\item\label{rule:two} \emph{At a 2 GB budget, take \BudgetTwoPick{}} [significant lead; frontier member].
It leads the best 1.5B file by \IsoThreeBOverOneFive{} points and stays on the frontier under every accounting, so the step from 1 GB to 2 GB trades energy for points.

\item\label{rule:anchor} \emph{Do not spend the next gigabyte on 7B-\qtwo} [repeated].
It solves \HeadSolvesB{} tasks against \HeadSolvesA{} for 3B-\qfour, and the 90\% CI limits any 3B-\qfour deficit to \HeadMaxDeficit{} points.
It needs \HeadJpsRatio{} times the energy per solve (\HeadJpsRatioSysLo{} to \HeadJpsRatioSysHi{} under whole-system accounting).
On HumanEval+, five alternating session pairs give a ratio of \TtwoRRailsThree{} against \TtwoAugustRRails{} in the main grid (Sec.~\ref{sec:rq3}).

\item\label{rule:fp} \emph{Skip \fp at 1.5B and 3B} [margin $\geq$40\%].
It stays off the frontier under every accounting (Table~\ref{tab:tests}, part E).

\item\label{rule:qfour} \emph{From 1.5B up, start at \qfour} [non-inferior at the post hoc margin].
Against the highest precision measured it costs \JpsOneFiveBQfour{} against \JpsOneFiveBFpsixteen{}~J per solve at 1.5B, \JpsThreeBQfour{} against \JpsThreeBFpsixteen{}~J at 3B and \JpsSevenBQfour{} against \JpsSevenBQeight{}~J at 7B, where the gap falls below 40\%.

\item\label{rule:beyond} \emph{At a 4 GB budget, choose by what you are budgeting} [leads unresolved].
For solved tasks, \BudgetFourPick{} leads: \BudgetFourPass\% against \BudgetTwoPass\% for the 2 GB pick.
For joules, stay on the 2 GB pick: \BudgetTwoJps{}~J against \BudgetFourJps{}~J per solve.

\item\label{rule:eight} \emph{At an 8 GB budget, move up only if the extra points are worth the energy} [lead unresolved].
The 7B files from \qfour to \qsix are not separated on solves, and \BudgetEightPick{} leads them by at most \SevenBQsixLeadMax{} points (90\% CI).
Against the 2 GB pick, the gain of \BudgetEightPick{} stays significant under Holm over all file pairs (p \AllPairsGainQsix{}), and those of 7B-\qfour and \BudgetFourPick{} do not (p \AllPairsGainQfour{} and \AllPairsGainQthree{}).

\item\label{rule:ties} \emph{Where a lead is unresolved or an energy ranking unreplicated, let the other axis decide.}
At 7B, \qfour costs \GapSevenBQfourOverQtwo\% more per solve than \qtwo [energy ranking unreplicated] and solves \SolvesSevenBQfour{} tasks against \SolvesSevenBQtwo{}.
Its lead over \qtwo stays significant under Holm over all file pairs (p \AllPairsSevenBQfourQtwo{}), so take \qfour when \ListedGBSevenBQfour{}~GB fits.
This also covers the 0.5B recipes at a 0.5 GB budget and two frontier calls too close to read: 0.5B-\qthree on by \MarginZeroFiveBQthreeOn\% and 7B-\qeight off by \MarginSevenBQeightOff\%.

\item\label{rule:smoke} \emph{Smoke-test downloads and leave headroom.}
The 3B-\qthree tag, non-functional in~\cite{patodiya2026p2}, fits the 2 GB budget at \ListedGBThreeBQthree{}~GB but solves \SolvesThreeBQthree{} of \NTasks{} tasks.
Per token it ranks \BrokenRankTok{}th of \NConfigs{}, and per solved task it ranks \BrokenRankJps{}th, at \BrokenJpsRatio{} times the energy per solve of its \qfour sibling.
Unusable \qtwo files of 3B Qwen2.5 models have been reported before~\cite{singh2026,synconfroute}.
Leave headroom above the listed size, because the runtime's buffers add \RuntimeOverheadMin{} to \RuntimeOverheadMax\% at load.
If free memory is the limit and not file size, those buffers, a lower bound on the runtime's allocation, move the 1, 2 and 4 GB picks in decimal GB to smaller files: \BudgetOnePickRuntime{}, \BudgetTwoPickRuntime{} and \BudgetFourPickRuntime{}.
Counted in GiB, that bound moves only the \RuntimeGibMovedBudget{}~GB pick.
\end{enumerate}

As of November 2026, the 1 GB and 2 GB picks are the files behind the registry's default tags \texttt{qwen2.5-coder:1.5b} and \texttt{qwen2.5-coder:3b}, and 7B-\qfour is the file behind \texttt{qwen2.5-coder:7b}.
From 1.5B up, then, a developer who takes a size's default tag stays on the frontier, and 7B-\qtwo, \fp and the 3B-\qthree tag are all off-default.

Our result is consistent with the four-bit optimum of Dettmers and Zettlemoyer~\cite{dettmers2023kbit}, for one code family with executed tests and energy.
The \qfour files we favor store \BpwOneFiveBQfour{}, \BpwThreeBQfour{} and \BpwSevenBQfour{} effective bits from 1.5B to 7B, so the shipped \qfour file is closer to five.
Registries could print effective bits per weight beside each tag.
A leaderboard column of energy per token should carry a solve count beside it, since a file can sit mid-table per token and still fail the tests.
A team can measure its own grid with the released harness and adapt the analysis script to rebuild the guide.
For a team that standardizes on this family on similar machines, the guide narrows trial downloads to a lookup by size budget that carries correctness and energy together.

For research, these results change what a study of quantized code models should print.
It should report effective bits beside tags, because a threshold stated in bits cannot be tested by tag.
It should price per solved task beside per token, because \TokFrontierHalfBExtraWord{} working 0.5B files that a per-token frontier keeps fall off the per-solve frontier, and Rule~\ref{rule:smoke} shows how far the two can part.
Under greedy decoding it should spend repeats on energy at decision steps, because in every repeat we ran outputs repeated byte for byte (Sec.~\ref{sec:repeats}) while energy and speed moved.

\section{Threats to Validity}\label{sec:threats}

\textbf{Energy measurement.}
Energy comes from Apple's IOReport estimates read by macmon~\cite{macmon,powermetrics}, with no external power meter.
Because the CPU, GPU and memory rails cover \RailsShareMin{} to \RailsShareMax\% of macmon's system power, we compare files on one machine as ratios and do not set absolute joules beside RAPL or NVML studies.
System power includes the display, which stayed on in the main grid and off in the repeat, so we do not compare system power between them.

\textbf{Sessions.}
Each grid file ran in one session per benchmark, back to back with no rest between sessions and in a fixed order by size and then recipe, so size and recipe align with time of day and with how long the machine had been under load.
Other applications were open during the main grid.
The paired repeat alternates the anchor files, which reduces the alignment between file and run order: with a cooldown gate before every block, its five-pair HumanEval+ ratio lies \TtwoBelowMainPct\% below the main-grid value.

\textbf{Paired repeat.}
The paired repeat covers the anchor pair on HumanEval+ in one night on AC power.
Thermal pressure stayed at Heavy for \TtwoHeavyMin{} to \TtwoHeavyMax\% of every block in both arms, so all \TtwoPairsFlagged{} pairs were flagged under the pre-specified rule and no unflagged subset exists.
We record this as telemetry and draw no thermal conclusion.
Browser and assistant apps were open, and automatic updates were on with none installed.
Pair 1 overlapped macOS media analysis, and Spotlight indexing ran just before it.
Without pair 1, a post hoc check gives \TtwoRRailsNoPairOne{} (\TtwoRRailsNoPairOneLo{} to \TtwoRRailsNoPairOneHi{}).
The pre-specified five-pair figure keeps pair 1, stays primary and is the lower of the two.

\textbf{Statistics.}
The \EquivMarginPP{}-point equivalence and non-inferiority margin was chosen after seeing the data, and the 40\% rule of Sec.~\ref{sec:stats} was set after the main grid.
J/solve intervals capture task sampling within the main-grid sessions.
Between-session variation comes from the paired repeat for the anchor pair on HumanEval+ in one night, and, as a rougher guide, from the two benchmark sessions of each file, which also differ in tasks.
Cross-size energy outside the anchor pair comes from single sessions.
The frontier uses point estimates, and under system power 1.5B-\qfour keeps its place by \OneGbSystemLeadWord{} solved task over 1.5B-\qsix, which lists outside the 1 GB budget.

\textbf{Idle baselines.}
The clean-idle baseline moves J/solve by \CleanIdleJpsMin{} to \CleanIdleJpsMax\% and leaves the anchor ratio at \HeadJpsRatioClean{}.
In the repeat, using the end-of-night idle or the main-grid idle instead of the start-of-night idle gives \TtwoRRailsEndIdle{} and \TtwoRRailsAugustIdle{}.

\textbf{Population.}
The study covers one code family, one runtime, one fanless machine, Python tasks, one zero-shot prompt and greedy decoding.
The grid covers \NConfigs{} of the 60 instruct recipes the registry lists from 0.5B to 7B.
It leaves out other recipes at these sizes, such as q4\_K\_S and q4\_0, and larger sizes, including the 14B files that list under 8 GB, so each pick is the best of the measured files.
Picks may move for other families, runtimes, cooling designs or languages.

\textbf{Registry files.}
Files are registry tags as pulled and identified by digest (3B-\qthree blob \BrokenBlob{}, manifest \BrokenManifest{}), so a re-pushed tag can be detected.

\textbf{Absolute scores.}
Our 7B-\qeight file scores \AnchorHeSevenBQeight\% on HumanEval+ and \AnchorMbppSevenBQeight\% on MBPP+, in line with Table 16 of the Qwen2.5-Coder report~\cite{hui2024qwen25coder}.

\section{Conclusion}\label{sec:conclusion}
A larger quantized code model is not automatically the better download, and a tag does not tell a developer which file that is.
The picks are \BudgetOnePick{} at 1 GB and \BudgetTwoPick{} at 2 GB, both on the frontier under every accounting.
At the step from 2 to 3 GB the larger model does not pay: 7B-\qtwo brings no significant gain in solves over 3B-\qfour, and the 90\% CI limits any 3B-\qfour deficit to \HeadMaxDeficit{} points.
The 7B file needs \HeadJpsRatio{} times the energy per solve (\HeadJpsRatioSysLo{} to \HeadJpsRatioSysHi{} on whole-system power).
At 8 GB, \BudgetEightPick{} buys a significant gain over the 2 GB pick at a higher energy price.
Registries could print effective bits beside each tag, and studies could price files per solved task beside per token.
The file a developer downloads is an energy decision as well as a correctness one, made once and with no change to code or hardware.

\section*{Data Availability}
The per-task correctness vectors, per-task energies, power logs, paired-repeat logs, file digests and the grid harness are in the artifact at \url{https://doi.org/10.5281/zenodo.24016482}.
The script \path{analysis/p1_numbers.py} writes every measured number in this paper from hash-checked inputs and the \NManifestsWord{} registry manifests it reads.
A fresh-directory run of the package on the study machine reproduced those numbers.

\section*{Acknowledgment}
The author conceived this work, chose the models, the quantization ladder, the benchmarks and the measurement protocol, directed the implementation, verified all reported results against the raw data and primary sources, and takes full responsibility for the content. AI assistance (Claude, Anthropic) was used to implement the measurement and analysis scripts and to draft the manuscript text under the author's direction.

\bibliographystyle{IEEEtran}
\bibliography{refs}

\end{document}